\documentstyle[12pt,psfig,epsf]{article}

\def\be{\begin{equation}}
\def\bea{\begin{eqnarray}}
\def\ee{\end{equation}}
\def\eea{\end{eqnarray}}

\def\rh{\rho}
\def\a{\alpha}

\def\s{\sigma}

\def\ra{\rangle}
\def\la{\langle}
\def\r{\right}
\def\l{\left}

\begin{document}
\title{The Modular Structure of Kauffman Networks}
\author{U. Bastolla$^{1,}$ $^2$ and G. Parisi$^1$}
\maketitle
\centerline{$^1$Dipartimento di Fisica, Universit\`a ``La Sapienza'', P.le Aldo
Moro 2, I-00185 Roma Italy}
\centerline{$^2$HLRZ, Forschungszentrum J\"ulich, D-52425 J\"ulich Germany}
\medskip
\centerline{Keywords: Disordered Systems, Genetic Regulatory Networks,}
\centerline{Random Boolean Networks, Cellular Automata}

\begin{abstract}
This is the second paper of a series of two about the structural
properties that influence the asymptotic dynamics of Random Boolean
Networks. Here we study the functionally independent clusters in
which the relevant elements, introduced and studied in our first paper
\cite{BP2}, are subdivided. We show that the phase transition in
Random Boolean Networks can also be described as a
percolation transition. The statistical properties of the clusters of
relevant elements (that we call {\it modules}) give an insight on the
scaling behavior of the attractors of the critical networks that,
according to Kauffman, have a biological analogy as a model of genetic
regulatory systems. 
\end{abstract}

\section{Introduction}
Kauffman networks are networks of randomly interconnected binary
elements under reciprocal regulation. Every node is controlled by
$K$ elements extracted at random in the set of the nodes. The boolean
control rules, which associate a binary value to every configuration of
the controlling elements, are also extracted at random at the
beginning and kept fixed during the evolution of the system.

The relevant parameter of the distribution of the control rules is $\rh$,
the probability that two different configurations of
the controlling elements produce a different output. 
For every value of the bare connectivity $K$ there is a critical value
of $\rh$, $\rh_c=1/K$, which separates a frozen phase (small $\rh$)
from a chaotic one (large $\rh$). According to Kauffman \cite{K69},
this model reproduces on its critical line some features of the
genetic regulatory systems acting in the cells.

In the preceding paper \cite{BP2} we defined the relevant elements of
Kauffman networks, which are the only elements that influence the
asymptotic dynamics,
and we measured their distribution. We refer to that paper for
the description of the properties of the model, of the distribution of
the relevant elements and of the relations between the number of
relevant elements and the properties of the limit cycles of the dynamics.
In the present paper we are going to describe how such elements are
spontaneously organized into dynamically disconnected clusters
that we will call {\it modules}.

It is known \cite{F} that some of the elements in the network evolve
to a constant state independent of the initial configuration and don't
take part to the asymptotic dynamics. The
relevant elements are the elements whose state is not
constant and that control at least one relevant element.
Thus every relevant element must be controlled by at least one relevant
element (eventually itself) and must control at least one of them
(eventually itself). It is then clear that loops of relevant elements
must exist. 

The set of relevant elements can thus be partitioned into different
clusters, that we call modules,
defined by the following equivalence relation: two relevant elements
belong to the same module if one of them controls the other one. It
is clear from this definition that elements in different modules
cannot communicate among each other after a transient time: thus the
dynamics induce spontaneously upon the network a clusterization of the
relevant elements. The number of the limit cycles and their lengths depend
only on this modular organization.

As expected, the statistical properties of the modules are very different
in the two dynamical regimes of the model. 

\begin{itemize}
\item In the frozen phase the number of relevant elements remains
  finite in the infinite size limit \cite{FK,BP2}. In this case, on
  the ground of the argument presented in \cite{BP2}, we
  expect that the modules are,
  in the infinite size limit, loops of effective connectivity
  exactly equal to 1 ({\it i.e.}, every relevant element is under the
  control of only one relevant element), and that the distribution of
  modules and, consequently, the number and the length of the attractors
  depend in this phase only on the parameter $c=K\rh<1$. 

\item In the chaotic phase the
  number of relevant elements is proportional to $N$, and the number
  of irrelevant elements tends to zero very fast as the bare
  connectivity increases (already for $K=3$ and $\rh=1/2$ the fraction
  of irrelevant elements is about $0.01$ for an infinite system). The
  set of the relevant elements is fully connected in the infinite size
  limit, where the number of modules tends to 1 also for $K=3$ and
  $\rh=1/2$,  as we shall see below. Nevertheless, in
  chaotic systems with low connectivity some signatures
  of a modular organization are present as a finite size effect.

\item At the border between the two phases, the critical line shows an
  interesting
  behavior. Here the number of relevant elements scales as  $\sqrt N$,
  {\it i.e.} the number of relevant elements increases with system
  size but they become more and more sparse. In this situation we
  expect that also the number of modules grows with system size, and
  that is actually what we observe in our simulations. The increase of
  the average number of modules is slow, apparently logarithmic
  in $N$, but it is responsible of important effects, such as the fact
  that the average weights of attraction basins tends to zero in
  the infinite size limit, or that the distributions of the length
  and the number of the limit cycles become broader and broader as $N$
  increases.
\end{itemize}

Thus the phase transition in Kauffman model is reminiscent of a
percolation transition. Below the threshold, the set of relevant
elements is divided into a finite number of finite clusters. Above, it
is completely connected. In critical systems, an infinite number
of clusters is present. In this case it is an open problem whether the
probability that all the relevant elements are connected tends to zero
or to a finite limit in the infinite size limit.

Derrida and Stauffer were the first ones to notice the analogy with
percolation theory for a lattice version of Kauffman model
\cite{DS}. The study of the modular organization shows that the analogy
with percolation theory remains pertinent also in the model with
random neighbors.
\vspace{0.5cm}

We start presenting our numerical results from the statistical
analysis of the divisors of cycle lengths (section 2), which allows the
simplest numerical investigation of the modular organization. 
Section 3 is about the direct measure of the distribution of the number and
size of modules, both in the chaotic phase and on the critical
line. In section 4 we show how the number of modules and the effective
connectivity depend on the number of relevant elements. A simple
argument predicts the qualitative behavior of these quantities in
agreement with numerical results.
Section 5 concerns the relation between the number of modules and the
length and number of the limit cycles, and section 6 deals
with the distribution of attraction basins weights, interpreting results that
we presented in \cite{BP1}.
An overall discussion concludes the paper.

\section{Divisors of cycle length}
\subsection{Analytic considerations in the frozen phase}
\begin{figure}
\centering
\epsfysize=15.0cm 
\epsfxsize=10.0cm 
\epsffile{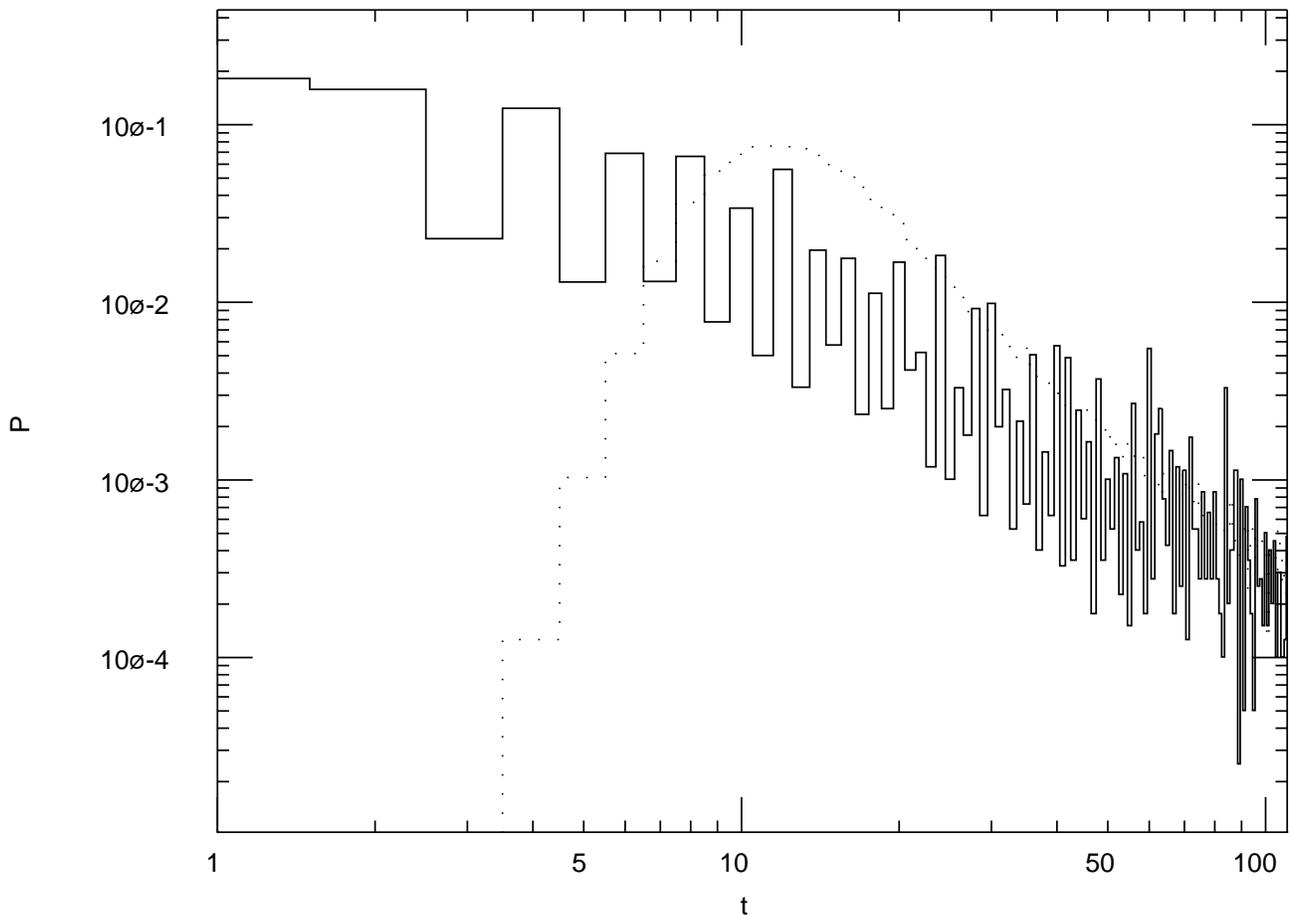}
\caption{\it Probability to find a cycle with period $t$ in a random network
with $K=2$, $\rh=1/2$ and $N=120$. The dotted line is the probability that
the transient time is equal to $t$.}
\label{fig_per2_120}
\end{figure}

The distributions of the length of the cycles in critical networks ($K=2$,
$\rh=1/2$) and supercritical  networks close to the phase transition
($K=3$, $\rh=1/2$) show an
oscillatory behavior: the cycles of even length are more frequent than
the odd ones (in other words, the closing probability is larger when a
cycle of even length is reached) \cite{BP0}. Figure
\ref{fig_per2_120} shows the period distribution for a network on the
critical line, with $K=4$ and $\rh=1/4$.

This is
what should be expected if the relevant elements are subdivided into
independent modules. In this case an attractor is the
composition of the local cycles of the different modules, and its
length is the least common multiple of the lengths of the local
cycles, so that if at least one of the modules has a local cycle of
even length also the global cycle will be even. Studying in more
detail the statistical properties of the numbers that are integer
divisors of cycle length it is thus possible to test this hypothesis and
to obtain qualitative informations about the distribution of the
modules, while the direct inspection of the
modules is numerically more cumbersome (with this method it is
enough to simulate one or two trajectories in every network,
while for the direct study of the relevant elements and of the modules
several hundreds of trajectories must be simulated).

We simulated two large systems in the chaotic phase and on the critical line,
generating thousands of sample networks and two trajectories on each of them.
For any small integer $l$ up to $l_{max}$ we measured the probability to 
find an attractor with length multiple of $l$, $P_d(l)$, and, only for
those networks where the two trajectories had reached different
attractors, we measured the conditional probability  that the length
of the second cycle is multiple of $l$ given that the length of the
first one is multiple of $l$ (we call this quantity $P_d(l|l)$). We
measured such quantities for different sets of networks, defined by the
condition that both cycles are shorter than a threshold length $L_0$,
and studied how the results vary as a function of $L_0$. As discussed
in \cite{BP2}, there is a positive correlation between the length of
the cycles and the number of relevant elements in a network, so that
we expect that the condition imposed on the length of the cycles
selects networks with less and less relevant elements as $L_0$ decreases. 

\subsection{Critical system}

We discuss separately the results obtained for the critical and for
the chaotic system, starting from the former one, which is a system with
2160 elements and parameters $K=4$ and $\rh=\rh_c=1/4$.

In this case the modular organization is very evident. In the case of
a Random Map, $P_d(l)$ is proportional to $1/l$. In the present case,
instead, we have to distinguish between values of $l$ which are prime
numbers and values which are not prime. For $l$ prime the probability
$P_d(l)$ has initially a value larger than $1/l$ (for  instance,
$P_d(2)=0.87$ when cycles shorter than 512 are selected), but then
decreases faster than $l^{-1}$. The decrease become faster when
shorter cycles are selected. This fact is consistent with the hypothesis
of the modular organization, and, within this interpretation, shows
that the probability to find a module with local cycle larger than $l$
decays faster than $1/l$, and even faster if the number of relevant
elements is small.

\begin{figure}
\centering
\epsfysize=15.0cm 
\epsfxsize=10.0cm 
\epsffile{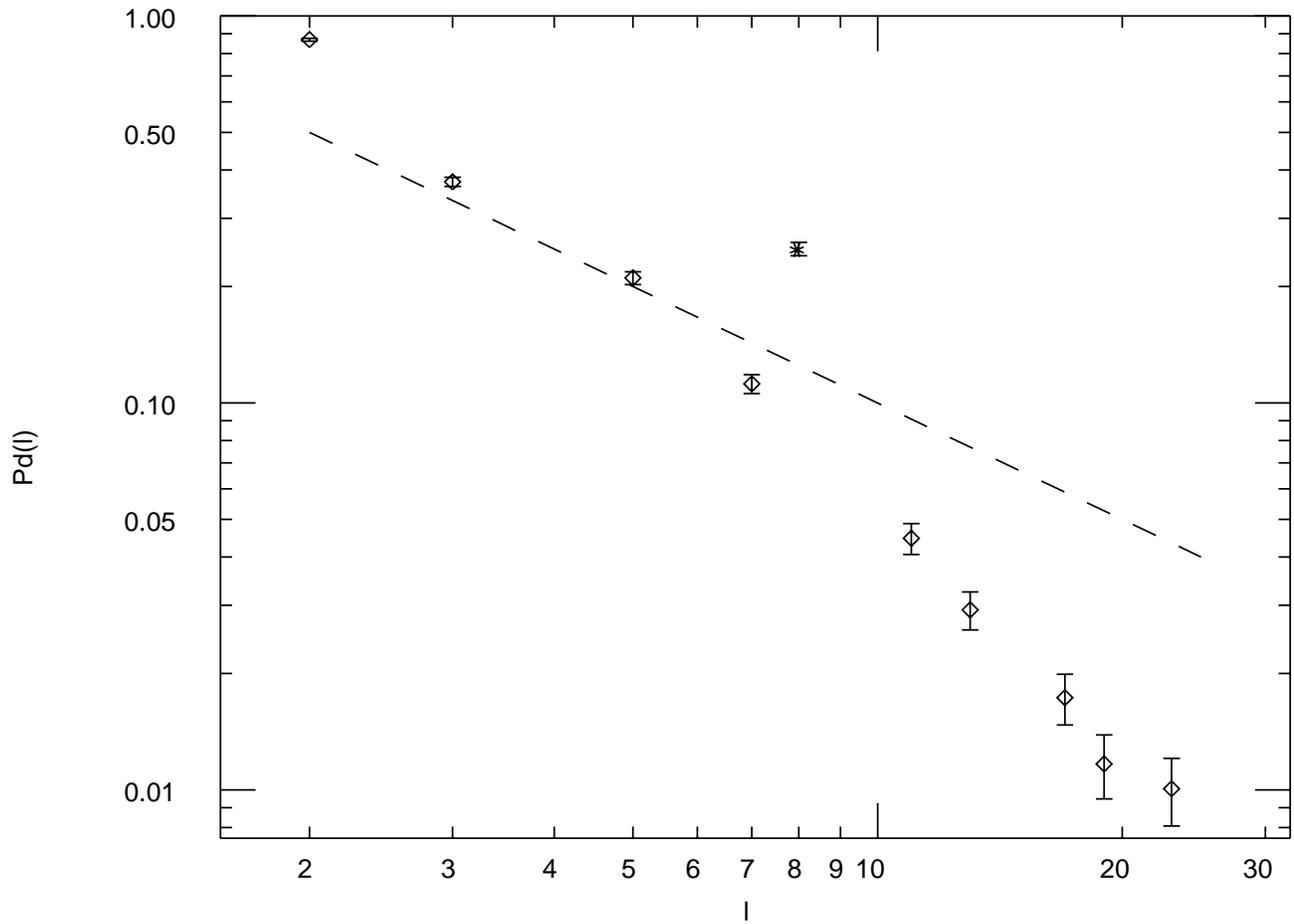}
\caption{\it Probability to find a cycle of length multiple of $l$, where
$l$ is a prime number, in critical networks with $K=4$, $\rh=1/4$ and $N=2160$.
The dashed line represents $1/l$. The isolated point above that line shows the
same probability for $l=8$.}
\label{fig_div4}
\end{figure}

If $l$ is not prime, on the other hand, $P_d(l)$ is larger than for
the prime numbers nearby, and it is usually close to the
product of the probability  of its prime factors (see figure
\ref{fig_div4}). This is a strong hint to interpret these prime
factors as the length of the local cycles of individual modules

\begin{figure}
\centering
\epsfysize=15.0cm 
\epsfxsize=10.0cm 
\epsffile{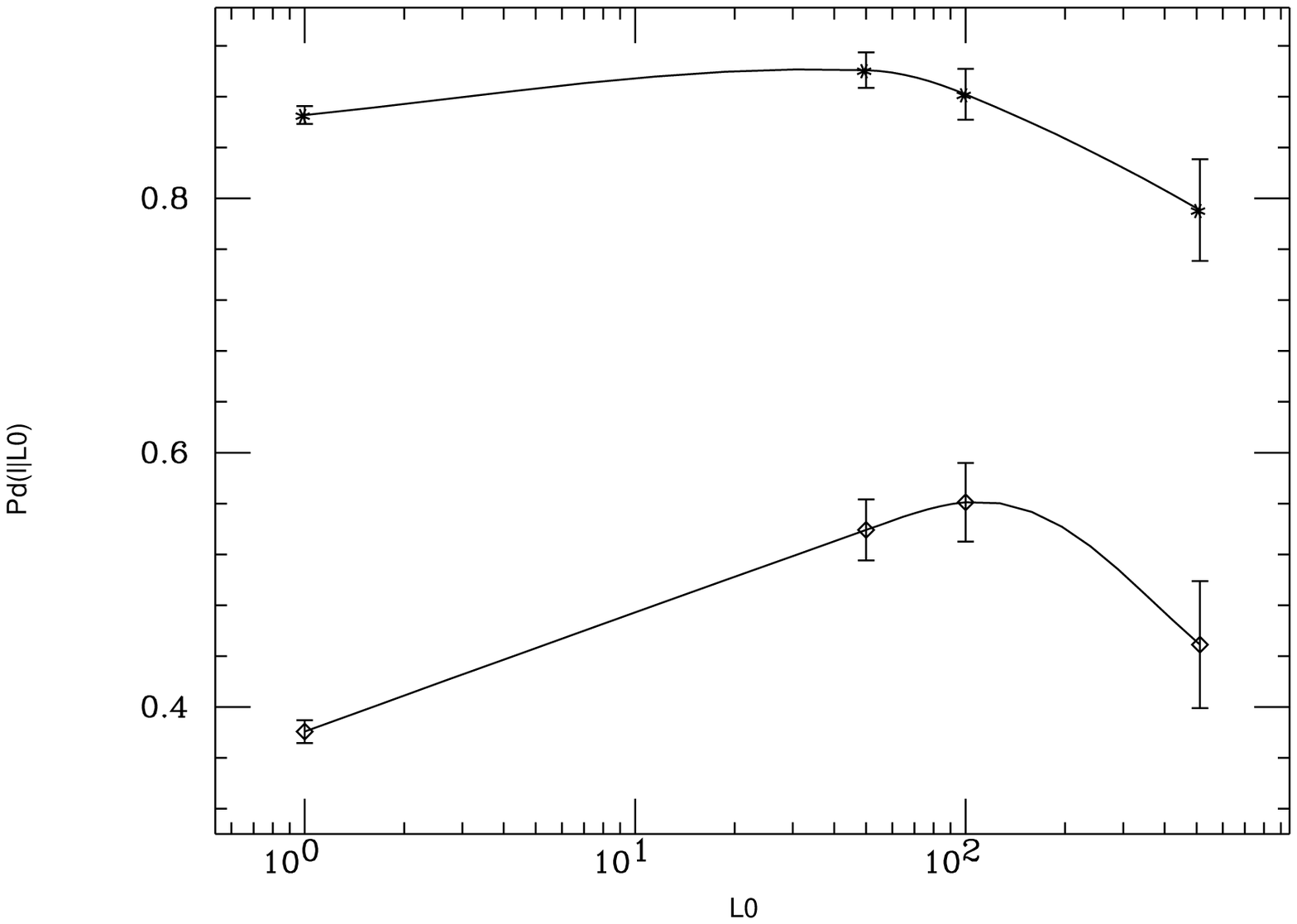}
\caption{\it Probability to find a cycle of length multiple of 2 (*) and
3 ($\diamondsuit$) restricted to the set of cycles shorter than $L_0$, as
a function of $L_0$, in critical networks with $K=4$, $\rh=1/4$ and $N=2160$.}
\label{fig_divl4}
\end{figure}

The effect of changing the threshold on cycle length, $L_0$, is shown
in figure \ref{fig_divl4}. For a fixed $l$, the probability to find a
cycle whose length is a multiple of $l$ increases, reach a maximum
value and then decreases. Since the abscissa can be related to an
increasing number of relevant elements, a possible interpretation of
this behavior is that the number of modules first increases with the
number of relevant elements, then reaches a maximum value and
ultimately decreases. We will see in section 4 that this is just
what happens.

At last, the condition that we already found in the same network a
cycle multiple of $l$ largely improves the probability to find another
cycle multiple of $l$, consistently with the modular interpretation.
This probability decreases when we select networks characterized by
longer attractors (this can be interpreted as another indication of
the fact that the number of modules may decrease as the number of
relevant elements increases).

%\begin{figure}
%\centering
%\epsfysize=15.0cm 
%\epsfxsize=10.0cm 
%\epsffile{divcond4.ps}
%\caption{\it  ($\diamondsuit$): Probability that the second cycle
%detected has a length multiple of $l$ with the condition that also
%the first cycle has it. (*): The same thing, but without imposing
%conditions on the first period. Both data are revered to the set of
%cycles longer than 50. $K=4$, $\rh=1/4$ and $N=2160$.}
%\label{fig_divcond}
%\end{figure} 

\subsection{Chaotic system}

The network that we examined is close to the critical phase: $K=3$,
$\rh=1/2$, $N=100$. This is a very large size concerning the
duration of the simulation, since in the chaotic phase the length of
the cycles increases exponentially with system size, but it still
shows important finite size effects. We think that the signs of the
modular organization that we observed can disappear in an infinite
system, since this signs decrease when we select networks
with longer attractors, {\it i.e.} with more relevant elements, and in
an infinite system the
fraction of relevant elements tends to $0.99$. A direct study of the
modules for smaller systems shows in fact that the average number of
modules tends to 1 in this limit (see next section).

Also in this case, however, the probability to find a global period
which is multiple of $l$ decreases faster than $1/l$ when $l$ is
a prime number, but the difference from $1/l$ is smaller than in
the previous case (in networks with attractors shorter than 50 steps,
its value is $0.70$ for $l=2$ and $0.29$ for $l=3$). Also this
probability is larger when $l$ may be decomposed into prime
factors. However this features tend to disappear when larger cycles
are selected, and $P_d(l)$ approaches $1/l$ when the threshold $L_0$
increases, thus indicating that the modular organization tends to
disappear when the number of relevant elements increases.

%(figure \ref{fig_divl3})

%\begin{figure}
%\centering
%\epsfysize=15.0cm 
%\epsfxsize=10.0cm 
%\epsffile{div3.ps}
%\caption{\it Probability to find a cycle of length multiple of $l$, where
%$l$ is a prime number, in chaotic networks with $K=3$, $\rh=1/2$ and $N=100$.
%The dashed line represents $1/l$. The isolated point above that line shows the
%same probability for $l=4$.}
%\label{fig_div3}
%\end{figure} 

%\begin{figure}
%\centering
%\epsfysize=15.0cm 
%\epsfxsize=10.0cm 
%\epsffile{divl3.ps}
%\caption{\it Probability to find a cycle of length multiple of 2 (*) and
%3 ($\diamondsuit$) restricted to the set of cycles larger than $L_0$, as
%a function of $L_0$, in chaotic networks with $K=3$, $\rh=1/2$ and $N=100$.
%The dashed lines are $1/2$ and $1/3$ respectively.}
%\label{fig_divl3}
%\end{figure} 

Correlations between the divisors of different
cycles on the same network seem to be present (the conditional probability
$P_d(l|l)$ is larger than $P_d(l)$), but they are of little
significance (the difference between the two probabilities is, at
most, of the order of two or three standard deviations) and our simulations
can not even prove that such correlations exist. In any case the
correlations, if any, are much smaller than in the critical system,
and they appear to decrease when networks with longer attractors are selected.

\section{Distribution of modules}

In this section we deal with a direct study of the number of
modules. Before presenting the results that we obtained, however, we
shall describe briefly our algorithm, also in order to give an
operative definition of the modules.

The first step, as it is discussed in \cite{BP2}, consists in
identifying the stable elements. To this aim we simulate 300
trajectories for each of the networks that we generate. The stable
elements are those ones that reach the same stable state in all the
different runs. The set of the unstable elements is then reduced to
the relevant elements alone (the ones that control at least one
relevant element), and in this process the irrelevant connections are
eliminated (the connection between $a$ and $b$ is irrelevant if
the response function does not depend on the state of $a$ when all the stable
values of the stable elements have been substituted in it, {\it i.e.} if
the function reduced to the unstable arguments alone is a constant
function of the state of $a$). Now every relevant element receives a
different label, and the following procedure is iterated: if either
$a$ controls $b$ or $b$ controls $a$, and their labels are different,
their labels are set equal to the smaller one, and all the labels
equal to the one that is changed are set to the same value. We check
that no label is changed more than once in an iteration. When no label
changes anymore the iterations are interrupted. At this point to every
different label corresponds a different module, and we count their
number and their size.

\subsection{Chaotic systems}
We start to present results relative to the chaotic phase, where
the existence of a modular organization appears only as a finite size effect.
We investigated numerically systems not far from the critical line,
with $K=3$, $\rh=1/2$ and $N$ comprised between 20 and 75. For such
systems, the probability to generate a network with more than one
module tends to zero roughly as $1/N$.
%(figure \ref{fig_mod3}).

%\begin{figure}
%\centering
%\epsfysize=15.0cm 
%\epsfxsize=10.0cm 
%\epsffile{mod3.ps}
%\caption{\it Distribution of the number of modules in chaotic networks
%with $K=3$ and $\rh=1/2$. $N$ is equal respectively to 20 (X), 40
%($\diamondsuit$), 60 (squares) and 75 (*).}
%\label{fig_mod3}
%\end{figure} 

Comparing this result with the analysis of the divisors of cycle length
(section 2) we see that the modular organization is not the only
explanation for the high frequency of even periods, which would be
smaller if it were due only to the presence of different
modules. It is possible however that some other structures
generalizing the notion of module can
give the same effect and are present with larger probability.
For example, we can define in-going and outgoing submodules as sets of
relevant elements that do not, respectively, receive from or send to the
outside any relevant signals. The length of a local cycle is then multiple of
the period of all the in-going submodules, and the situation described
in the previous section may occur because of the submodules.
Another structure that could play a role is that of cycle-depending
modules. Globally stable elements are defined as the elements that
reach the same fixed state independently on the initial
configuration. The elements stable on a cycle are a subset of those,
and are defined as the elements whose value does not change along the cycle.
Thus the network can be decomposed into cycle-depending
modules that do not communicate between each other when the system is on the
attractor. Once again, the period of the attractor is the least common
multiple of the periods of all cycles of the partial modules that
compose it. Or it may be that the statistical properties of the
divisors of cycle length do not depend on the presence of modules.

However, it is clear that the modular organization tends
to disappear in the infinite size limit.

\subsection{Critical systems}
\begin{figure}
\centerline{\psfig{figure=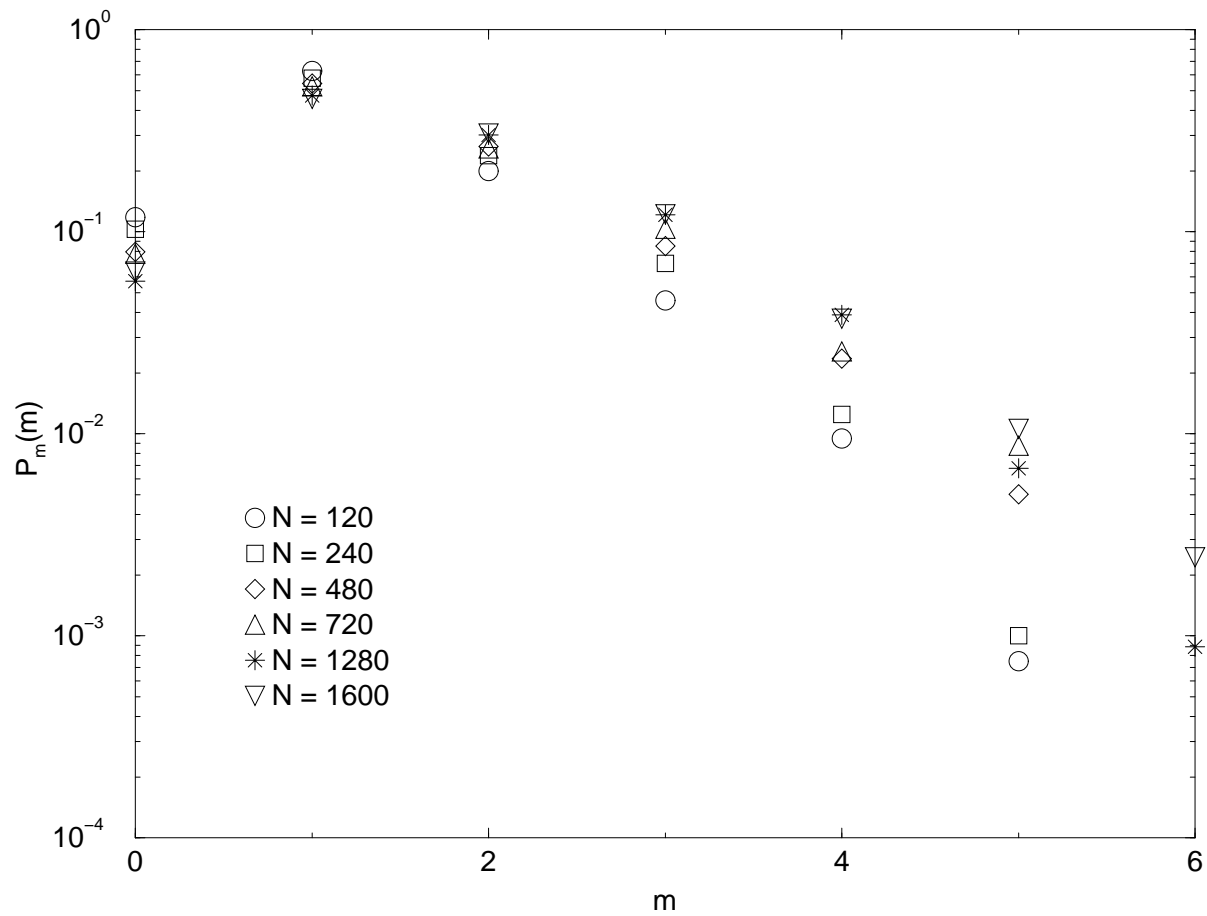,height=12cm,angle=-90}}

\caption{\it Distribution of the number of modules in a random network at the
critical point $K=4$ and $\rh=1/4$. $N$ is ranging from 120 to 1600.}
\label{fig_mod4}
\end{figure}

The distributions of the number of modules
in critical systems with $K=4$ and $\rh=1/4$ have a maximum
corresponding to a single module and then decrease roughly
exponentially, but with a rate that becomes slower and slower as
system size increases (figure \ref{fig_mod4}). The probability to find
a single module, or no module at all (this means that all the elements
are stable) decreases with system size, while the probabilities to find
2 or more modules increases. It is not clear whether the probability
of networks with only one module tends to zero or to a finite limit.
The average number of modules increases with system size. T, and the
best fit is logarithmic:
\be \l\la N_{mod}\r\ra\approx 0.40+0.39\log N. \ee

We prefer the logarithmic fit to a power law with a small exponent,
also compatible with our data, for
two reasons: first, because the inclusion of the largest $N$ value
modifies the fit parameters by 5 percent in the first case and by 10
percent in the second one, and second because at the ``critical point"
$K=1$, $\rh=1$ the analogy between the modules and the cycles of a
Random Map shows that the average number of modules grows
logarithmically with system size in this case \cite{FK}. This point is
not a real critical point since it is not at the border between two
phases (with $K=1$ only the frozen phase is possible), but still the
analogy with it can give useful indications.

\begin{figure}
\centering
\epsfysize=15.0cm 
\epsfxsize=10.0cm 
\epsffile{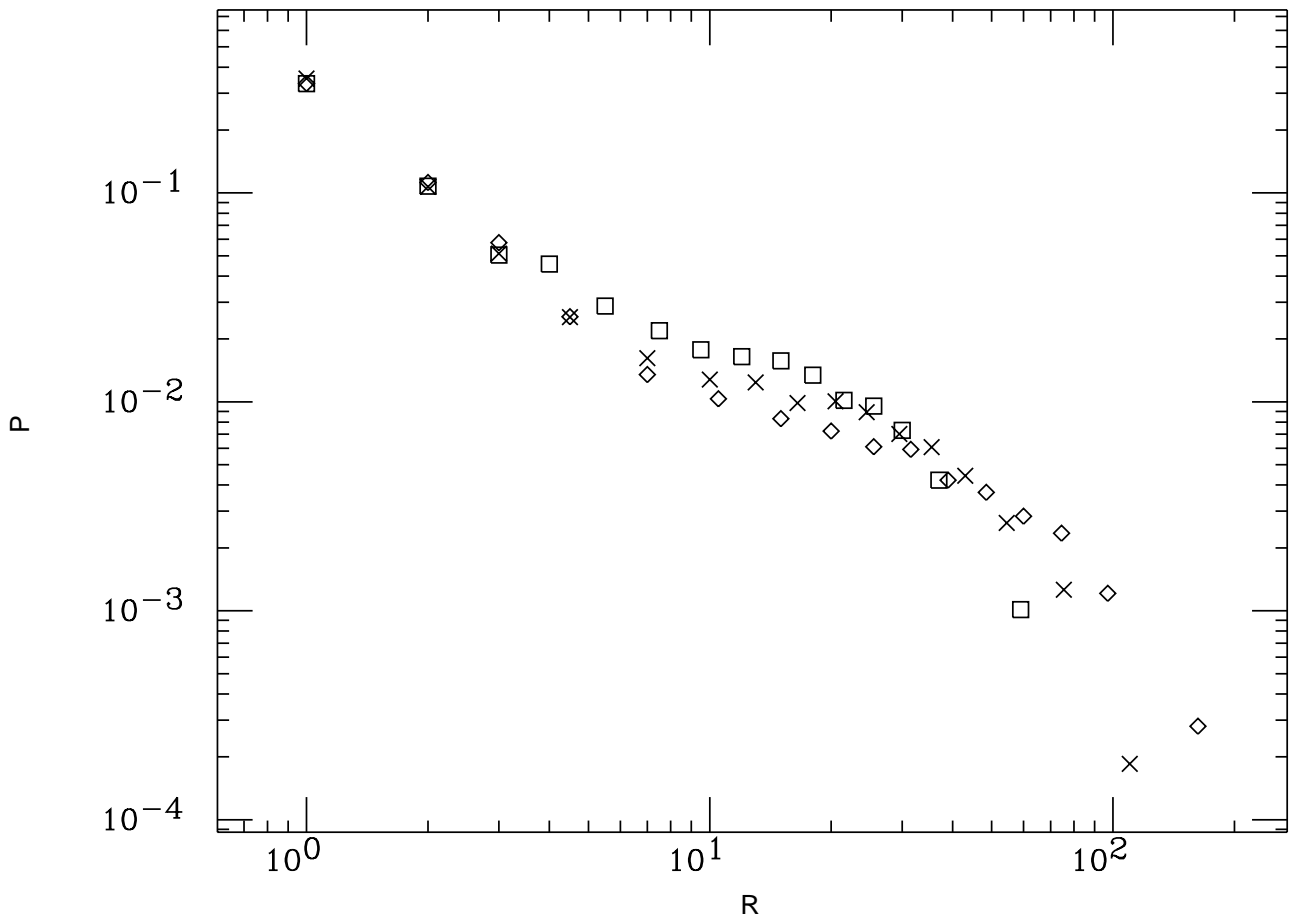}
\caption{\it Distribution of the number of modules with $R$ relevant elements,
at the critical point $K=4$, $\rh=1/4$; $N$ is respectively equal to
120 (squares), 240 (X) and 480 ($\diamondsuit$).}
\label{fig_mods4}
\end{figure}

Most of the modules have only one element, and the distribution of the
size of the modules decreases monotonically. The probability to find a
module with 1, 2 or 3 elements depends very little on the system size $N$. On
the other hand, this probability increases with $N$ for modules of
large size (figure \ref{fig_mods4}), consistently with what we saw in
\cite{BP2}: the number of relevant elements, which is the sum of the
sizes of all the modules, scales as $\sqrt N$ on the critical line.

\section{Effective connectivity}
In this section we examine how the number of modules depends on the number of
relevant elements. A useful notion to understand this relation is the
effective connectivity of the network.  Also in this case the
simulations were done with critical networks with $K=4$ nd $\rh=1/4$.

\begin{figure}
\centering
\epsfysize=15.0cm 
\epsfxsize=10.0cm 
\epsffile{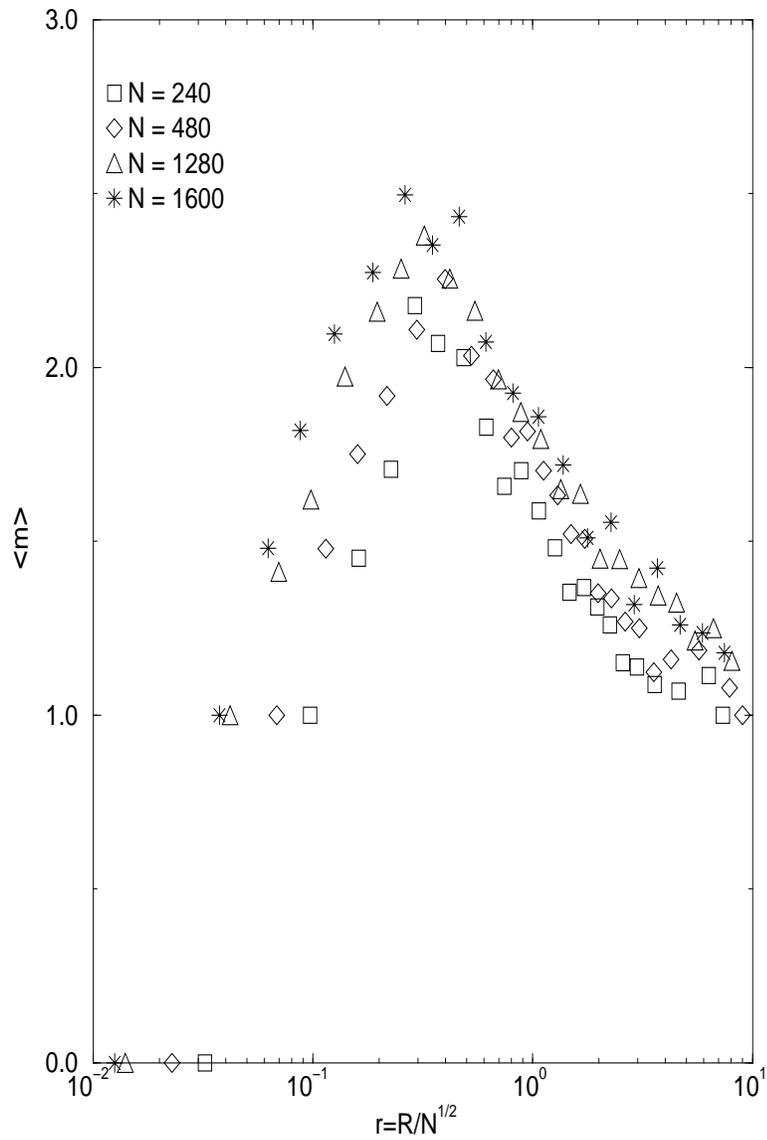}
\caption{\it Average number of modules in networks with $R$ relevant nodes as a
function of $r=R/ N^{1/2}$ at the critical point $K=4$, $\rh=1/4$.}
\label{fig_modril}
\end{figure}

We plot in fig. \ref{fig_modril} the average number of modules as a
function of the number of relevant elements in the network, $R$, for
systems of different sizes. This quantity shows a non monotonic behavior:
the number of modules increases with $R$ at the beginning, then it
reaches a maximum and starts to decrease towards the value 1.

\begin{figure}
\centering
\epsfysize=15.0cm 
\epsfxsize=10.0cm 
\epsffile{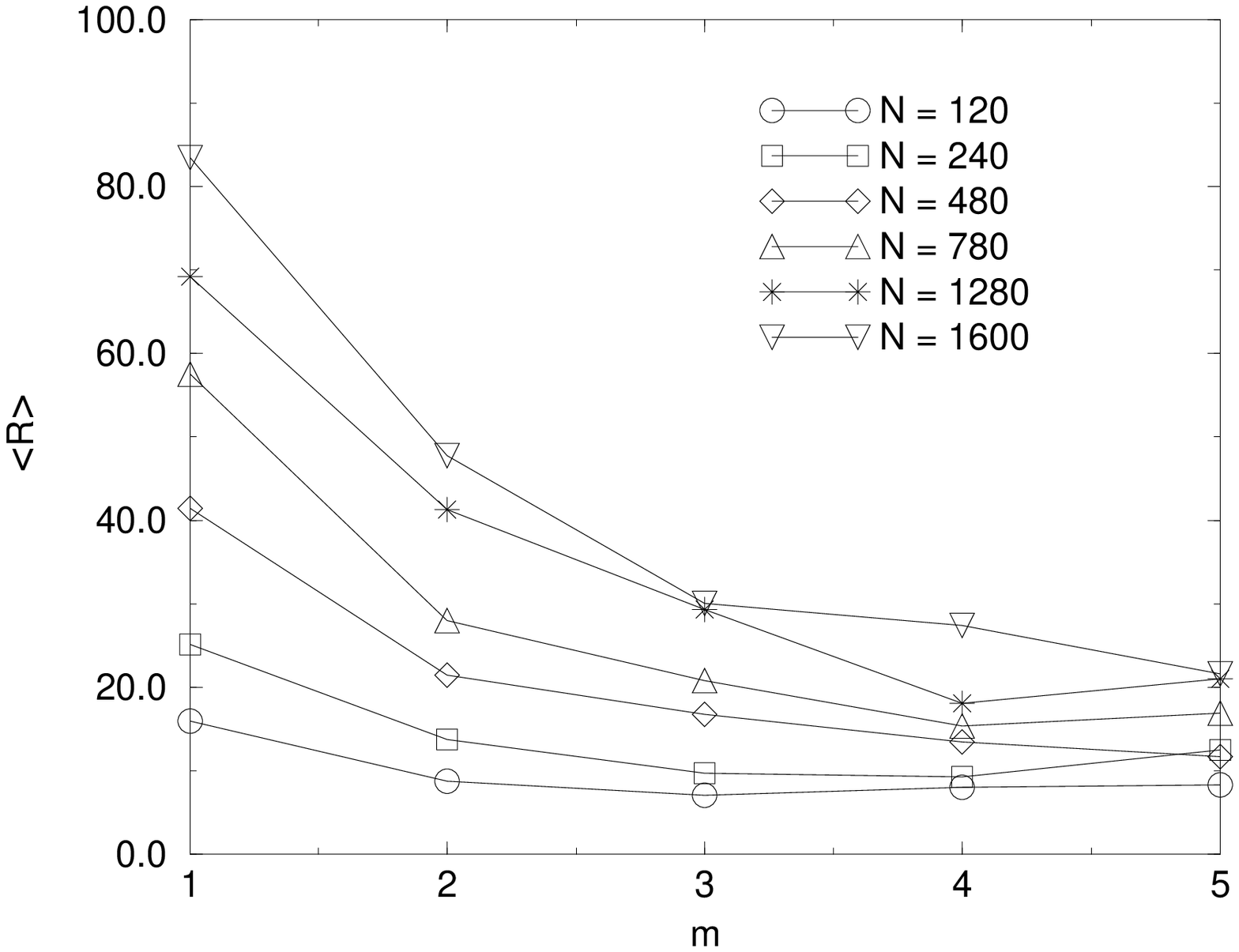}
\caption{\it Average number of relevant elements in networks with $m$ modules
at the critical point $K=4$, $\rh=1/4$.}
\label{fig_rilmod}
\end{figure}

The plot of the average number of relevant elements as a function of
the number of modules (figure \ref{fig_rilmod}) contains a similar
information. Contrarily to the naive expectation, the largest number of
relevant elements is attained in networks with only one module. It
seems that then the curve reaches a minimum and
starts to increase, but we have the networks with such a large
number of modules are very rare and we can not make any significant
statement about this point.

It is not difficult to explain qualitatively this behavior.  At this
end we use the same argument that we used in \cite{BP2} to
estimate the scale of the number of relevant elements. We suppose to
add a new element to a network with $R$ relevant element. If the new
added element happens to be relevant, three situations are possible:

\begin{enumerate}
\item The new relevant element is added to a previously existing
  module and the number of modules does not change;

\item The new relevant element constitutes a new module itself and the
  number of modules increases by one unit;

\item The new relevant element connects two modules and the number of
  modules decreases by one unit.
\end{enumerate}

When $N$ is large all other possibilities can be neglected. The
situation is trivial in the frozen phase, where an asymptotic (for
large $N$) distribution of the relevant elements is reached and all
the three cases above have negligible probability, and in the chaotic
phase, where only one module is present in systems of large size, but
it is very interesting on the critical line, since the properties of
the attractors vary according to which of the three situations is the
typical one.

To understand how the number of modules depends on $R$ on the critical
line, we have to
compare case 2 and case 3. If, for a given value of $R$, the second
case prevails, the number of modules increases with $R$. In the other
case, it decreases. The probability of case 2 should not depend on
$R$, while the probability of case 3 is proportional to $\sum_i
n_in_j$, where $n_i$ is the size of module $i$ and the sum runs over
different modules. When $R$ increases, we should expect this quantity
to increase so that case 3 prevails. The quantity that allows to
understand quantitatively the situation is the average rescaled size of a
module, defined as

\be S_2=\l\la \sum_i n_i^2/\l(\sum_i n_i\r)^2\r\ra . \ee

If this quantity is large (close to 1) case 2 prevails also for large
values of $R=\sum_i n_i$, otherwise case 3 is the prevailing one.

This argument is supported by the study of the average number of
relevant connections\footnote{In this context, we say that element $a$
  has a connection with element $b$ only if $a$ is controlled by $b$,
  and not if $b$ is controlled by $a$.}
 in the network, that we called the {\it effective
connectivity} of the network. In critical networks, this quantity is much
smaller than the bare connectivity, $K$, and tends to 1 as a power law when
$N$ increases. We are interested to the behavior of $K_{eff}$ as a
function of the number of relevant elements in the network (in other
words, $K_{eff}(R)$ is a conditioned average, restricted to the set of
networks with exactly $R$ relevant elements).

\begin{figure}
\centering
\epsfysize=15.0cm 
\epsfxsize=10.0cm 
\epsffile{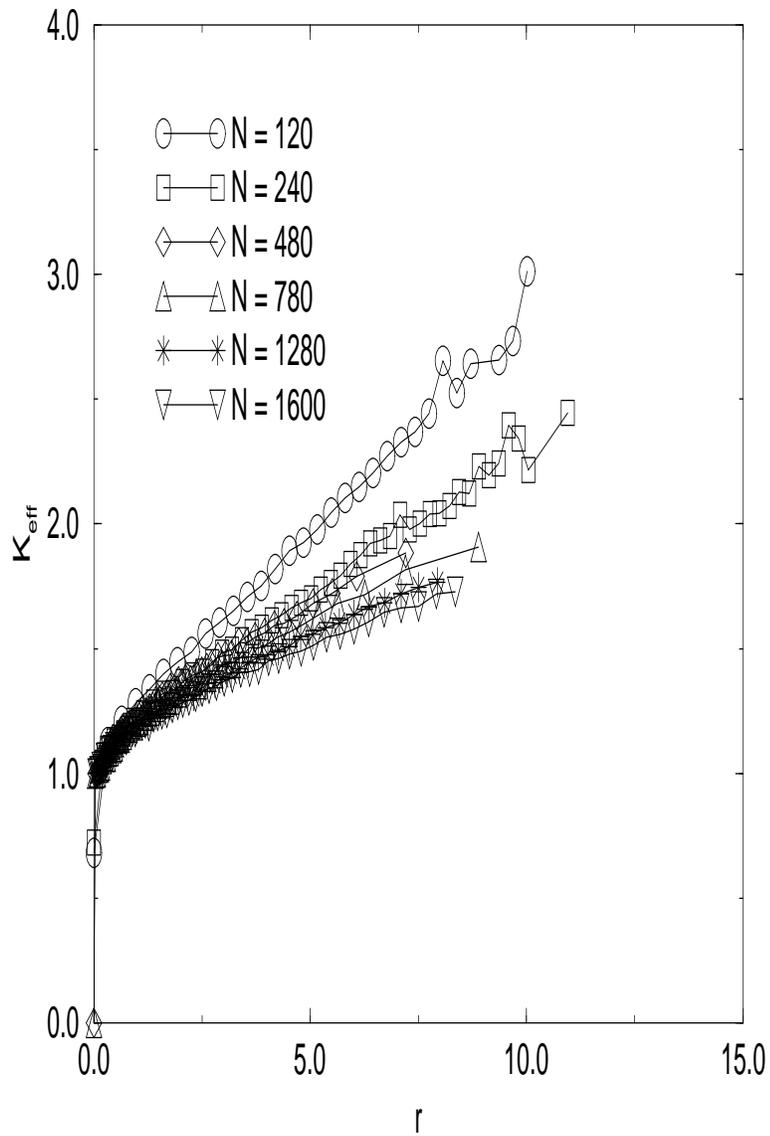}
\caption{\it Effective connectivity in networks with $R$ relevant nodes as a
function of $r=R/N^{1/2}$ at the critical point $K=4$, $\rh=1/4$.}
\label{fig_keff}
\end{figure}

For large $R$ this function is asymptotic to a straight line with
intercept close to 1 (see figure \ref{fig_keff})\footnote{
Of course, $K_{eff}$
should saturate to a value smaller than $K$ when all elements are
relevant, but such networks occur with negligible probability on the
critical line.}. It holds:

\be K_{eff}(R) \approx a(N)R+b(N), \ee
where the parameter $a$ decreases as a power law: $a(N)\approx 1.7N^{-0.93}$,
and $b(N)$ tends to 1 from above (every relevant element must have at
least one relevant connection, so that $K_{eff}$ can not be less than 1).
 
This is very close to what we expect on the basis of the above
argument \cite{BP2}. We expect that the conditional probability that the
new element has two relevant connections, given that it is a relevant
element, is proportional to $R/N^2$. This is a quantity of order $1/N$, so this
argument predicts $a(N)\propto 1/N$, to be compared with our best numerical fit
$a(N)\propto N^{-0.93}$. 
From this argument, we would expect also that $K_{eff}=b(N)+\la R\ra
a(N)$ is equal to 1 plus a term vanishing as $1/\sqrt N$. It turns out,
however, that the correction decreases as $N^{-0.3}$: this discrepancy
is due to the term $b(N)$, and we don't have a good explanation for it.

The behavior of the effective connectivity and that of the number of
modules as a function of $R$ show the existence of a crossover between
two situations reminiscent of the two phases as the number of
relevant elements changes along the critical line. For small $R$ we
find networks with small effective connectivity: most of the modules
are loops where each relevant element has only one connection, as it
is typical in the frozen phase. To the opposite side, where $R$ is
large, the networks are much more connected and there are few modules.
This situation is similar to what happens in chaotic networks, but on
the critical line only a finite number of elements have more than 1
connection in the infinite size limit.

We thus expect that the properties of the attractors change widely
with the number of relevant elements, from a situation reminiscent of
the frozen phase to a situation reminiscent of the chaotic phase. We
dedicate the rest of this paper to describe how the properties of the
attractors are influenced by the properties of the modules.

\section{Attractors and modules}
\subsection{Simply connected modules}
We saw in the previous section that the effective connectivity of
critical and frozen systems tends to 1 in the infinite size
limit. This means that most of the elements receive and send only
one relevant signal. Thus simply connected modules play an important role in
critical networks, and are found with probability one in the frozen phase.
Such modules were analytically studied by Flyvbjerg and Kjaer in the
case of $K=1$ networks \cite{FK}. Here we recall for completeness their 
properties, adding some more information.

A simply connected module is a loop of relevant elements $i=1,\cdots l$,
such that the node $i-1$ controls node $i$ through the Boolean function
$f_i(\s)$ (we assume periodic boundary conditions). There are only two
kinds of non-constant Boolean functions of one Boolean argument: tautology
($f(\s)=\s$) and contradiction ($f(\s)=\overline{\s}$).

The number and length of the cycles of such modules do not depend on
the details of the distribution of the Boolean functions, but only on
the parity of the number of negative functions \cite{FK}.  All the
$2^l$ configurations of the module belong to some cycle of 
length equal to an integer divisor of $l$ (if the number of negative functions 
is even) or of $2l$ (if it is odd): in fact, after at most $l$ steps in the 
first case, and $2l$ steps in the second one, every initial
configuration is reproduced.  This is the first important property:
the length of the cycles increases only linearly with the size of the module.

The number $g_l(m)$ of cycles of length $m$ in a module of size $l$ is
easily found. First, it is easy to see that $g_l(m)$ does not depend on
$l$. In fact, let us assume an even number of negative functions. A module made
out of $l$ elements may be thought of as the repetition of $k$ modules
made of $m$
elements, as $m$ must be an integer divisor of $l=km$. But for all the initial
configurations whose periodicity is $m$ the dynamics is independent
on the value of $k$; on the other hand, the configurations that are not
periodic in space can not be reproduced after $m$ steps. For an odd number of
negative functions the argument is exactly the same, but the $k$
modules of size $m$ can be equal only if $k$ is odd. Thus $g_l(m)$
depends only on $m$ and on the parity of the number of negative
functions, that we shall denote by a superscript: $(0)$ to indicate
even, and $(1)$ to indicate odd. It holds

\be \sum_{\{m\mid l/m \, integer\}} m g^{(0)}(m)=2^l, \ee
\be \sum_{\{m\mid l/m\, odd\}} 2m g^{(0)}(2m)=2^l, \ee
whence it can be seen that $g^{(0)}(l)$ is asymptotic to $2^l/l$ and
$g^{(1)}(2l)$ is asymptotic to $2^{(l-1)}/l$. In the large $l$ limit,
most of the cycles have the maximal length.

\subsection{Global attractors}
A global attractor is made out of the composition of the limit cycles
of all the modules. Its period is thus the least common multiple of the
periods of the cycles that compose it. The number of attractors is
larger than or equal to the product of the number of cycles in the different
modules. In fact the number of attractors that can be formed with a
set of limit cycles, each one taken from a different module, is equal
to the maximum common divisor of the periods of the cycles.
 
%Thus the knowledge of the number, size and connectivity of the modules
%would give us every information about the number and the length of the
%attractors of the system. This is specially interesting for critical
%networks, for which there is not a satisfactory theoretical
%description of the properties of the attractors and numerical results
%do not give clear indications about their behavior asymptotically with
%system size \cite{BP1}.

Even in the absence of an analytic description of the statistical
properties of the modules, the study of the modular structure allows
us to better interpret the numerical results about the distributions
of the number and of the length of the attractors in critical networks
\cite{BP1} and sheds light on the large size behavior of these
quantities. In the following we will concentrate our discussion on
critical networks.

First, we note that the two extremal situations of very high and very
low connectivity, which are reminiscent respectively of the chaotic
and of the frozen phase, are expected to be very different also from
the point of view of the properties of the attractors. 

In high connectivity critical networks the attractors are reminiscent
of the chaotic phase (we have to remember, nevertheless, that for
critical networks $K_{eff}$
tends to 1 in the infinite size limit, so that high connectivity means
that there is a large but finite number of elements with two
connections. Such networks are characterized, as it is discussed
above, by few modules of large size).
We observed in fact that the average length of the cycles increases
exponentially with the number of relevant elements for large $R$, as
it happens in the chaotic phase. Thus in such networks the average
length of the cycles should increase as a stretched exponential of
system size. This result is consistent with the fact that the
distribution of the length of the cycles seems to have an effective
scale increasing as a stretched exponential \cite{BP1}. 
Similarly, we expect that the number of attractors in such networks is
not very large (in the chaotic phase the number of attractors
increases linearly with the number of relevant elements), but we don't
have data to compare to this prediction.

In low connectivity critical networks there are several modules with very
few elements. Most of the modules are just loops with
effective connectivity exactly equal to 1, so that the length of the
local periods is proportional to the size of the loop and the number
of cycles increases almost exponentially with it. Thus, in this
situation we expect to find a very large number of attractors,
exponentially increasing with the number of relevant elements. This
means that in such networks the number of cycles should increase
approximately as a
stretched exponential of system size. The distribution of the number
of attractors is consistent with this prediction, since it has an
effective scale increasing as a stretched exponential of system size. 
On the other hand, we expect to find in such networks short
cycles: the typical length should be the minimal common multiple of
the size of the modules, and it should increase as the square root of the
number of elements, as far as the number of modules is not very
large. But, as $N$ increases, this scaling should be lost
and substituted by a faster increase.

It can thus be
understood why both the distributions of the length of the cycles and
the distribution of their number become broader and broader when
system size increases: in the ensemble of critical networks quite
different situations coexist. It seems also that networks
characterized by a small number of short attractors (meaning by this that
the number and the length increase not faster than a power law of $N$)
can be found only if the number of relevant elements is very small, thus with
vanishing probability as $N$ increases. However, this qualitative picture still
requires more accurate further investigations to be confirmed or rejected.

\section{Attraction basins}
While the number and the length of the attractors
depend only on the number and the size of the modules, this is
not true for the size of the attraction basins.

In fact, when we extract at random an initial configuration, the
elements that are going to become stable are not yet stable and the
modules are not yet formed (the case $K=1$ is an exception, since in
this case the modules are present from the beginning), since the elements
of different modules can still
communicate. Thus the attraction basins of two critical networks with
the same modular structure but with different bare connectivity don't
need to be identical. In fact, we expect that the attraction basins
for critical systems are larger the larger is $K$, since then
$\rh_c=1/K$ is smaller and a large number of the input configurations
produce the same state. In the following, we will be mainly dealing with
critical systems, unless otherwise stated.

The simulations of critical systems with $K=2$ and $K=4$ confirm that,
for a given system size, the attraction basins are on the average
larger for $K=4$ than for $K=2$ \cite{BP1}, but show also that the
main characteristics of the distribution of the weight of the
attraction basins are the same in both cases. The weight of the
attraction basin $\a$, $W_\a$, is defined as the fraction of
configurations that evolve towards attractor $\a$. This quantity has
to be averaged over all the attractors and over several
realizations of the dynamical rules. Following \cite{DF1} we use the notation

\be \l\la Y_2\r\ra=\sum_\a \l\la W^2_\a\r\ra . \ee
(the angular brackets represent the average over the disorder).

For both the $K$ values that we examined, $\la Y_2\ra$ vanishes as system size
increases, and in both cases our data are compatible with a power law
decay with the same exponent for the two parameters considered,
suggesting that this behavior is universal along the critical line.  
Moments of higher order of the weight of the attraction basins also
seem to vanish as a power law of system size, but with an exponent
which is not proportional to the order of the moment, so that a
typical weight of the attraction basins can not be defined and the
 distribution of the weights has multi-fractal properties \cite{PV}.

\begin{figure}
\centering
\epsfysize=15.0cm 
\epsfxsize=10.0cm 
\epsffile{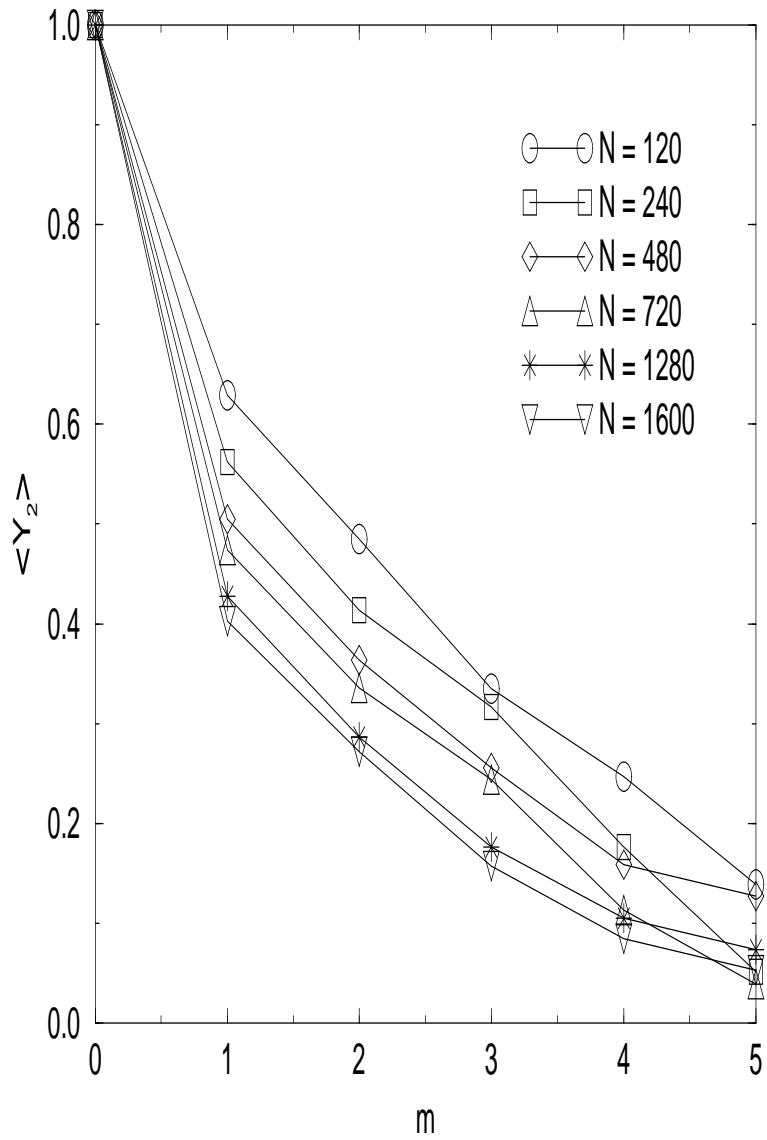}
\caption{\it  Conditional average of $Y_2=\sum_\a W_a^2$ in networks with
$m$ modules. $K=4$, $\rh=1/4$.}
\label{fig_ymod}
\end{figure}

The fact that the average weight of the attractor basins vanish in the
infinite size limit can be qualitatively explained considering the
modular structure of critical networks. We measured $\la Y_2\ra$
selecting only networks with a number $m$ of modules and we observed
that this conditional average is a decreasing function of $m$, as
figure \ref{fig_ymod} shows.

This behavior has a clear explanation in
networks with $K=1$, where the modules are independent
form the beginning. In this case, two configurations belong to the
same attraction basin if and only if all their components on different
modules belong to the attraction basins of the same local cycles. Since all
such components are extracted independently, the probability of such an event
is the product of the probabilities relative to individual modules and
vanishes as the number of modules increases. This reasoning can not be
applied to networks with connectivity 
larger than one, as we said previously, but the result that the average
value of $Y_2$ decreases with the number of modules remains valid also in
this case.

On the other hand, the average weight of the attraction basins has not a
monotonic behavior as a function of the number of relevant elements in the
network, $R$. This fact can be explained as a consequence of the fact that
the number of modules does not grow monotonically
with the number of relevant elements in the network. In the
low-connectivity regime the number of modules increases with $R$, thus
the weight of the attraction basins decreases. The contrary happens in
the high-connectivity regime.

\vspace{0.5cm}
These characteristics do not change in the chaotic phase.
Also for chaotic systems the average weight of the attraction basins 
decreases with the number of modules in the network, but in this case the
probability to find more than one module goes to zero in the infinite size
limit, and the effects of the modular organization disappear. In particular,
the average basin weight remains finite and its limit value appears to be
very close to the Random Map value $\la Y_2\ra=2/3$, as predicted by the
annealed approximation \cite{BP0}.

We observed in simulations reported in \cite{BP0} that in
chaotic systems not far from the critical line ($K=3$, $\rh=1/2$) the average
basin weight,$\la Y_2\ra$ , is not a monotonic function of system size:
at first it decreases with $N$, then attains a minimum value of about $0.59$
at $N=40$ and starts to increase, apparently toward the Random Map value
of 2/3. One could think that this not monotonic behavior is related to the
modular structure, supposing that its effects are not monotonic in system
size. This does not appear to be the case, as the same not monotonic behavior
as a function of system size can be observed restricting the sample to
networks with a single module (they are the big majority even for small sizes,
in the chaotic phase). Nevertheless, it is possible that the extension of
the concept of modules that we proposed in section 2 in order to fill the
gap between the observed distribution of modules and the effects of
factorization of cycle lengths can give a better comprehension of this
behavior.

\section{Discussion}

The dynamical transition taking place in Kauffman networks can be
characterized as a percolation transition in the set of the relevant
elements (the analogy with percolation  was already
proposed by Derrida and Stauffer with respect to Random boolean
networks on a lattice \cite{DS}).

In the frozen phase, the relevant elements are finite in
number also in the infinite size limit and they are divided into asymptotically
independent modules. In the chaotic phase all the relevant elements belong
to the same cluster, and their number is proportional to system size.
At the border between these two phases, the critical line is
characterized by a number of relevant elements increasing as $\sqrt N$
and a number of modules that grows with system size (apparently with a
logarithmic behavior).

Critical networks have a wide range of modular structures, ranging from
networks reminiscent of the frozen phase (with few relevant elements and
low effective connectivity and a number of modules that increases with
the number of relevant elements) to networks reminiscent of the
chaotic phase (with many relevant elements, larger effective
connectivity and a number of modules that decreases with
the number of relevant elements).

The modular organization can explain qualitatively -- and, maybe, also
quantitatively -- the main properties of the attractors of critical
networks. These attractors are the composition of the undecomposable cycles of
the elements of the single modules, so that their length is the
minimal common multiple of the lengths of these cycles and their
number is the product of the numbers of these cycles. One expects then
that networks closer to the frozen phase, whose modules have effective
connectivity very close to 1, are characterized by a large
number of attractors (exponentially increasing with $R$, the number of
relevant elements) whose length increases as $R$ or faster. In
networks close to the chaotic phase, on the other hand, one expects to
find less attractors (for the same value of $R$), but longer ones. This is
in agreement with the observed distributions of the number and of the
length of the attractors, that become broader and broader as system
size increases.

It seems to us that these results imply that the
typical values of the number and of the length of the attractors
increase asymptotically faster than any power law. In simulations of
small systems, Kauffman \cite{K69} observed that this quantities
behave as $\sqrt N$. Simulating larger critical systems \cite{BP1}, we
observed that the distributions of these quantities are compatible
with two effective scales, one increasing as $\sqrt N$ and another
one, for rare but not vanishingly rare networks, increasing as a
stretched exponential of $N$. This observation is consistent with the
conclusions that can be drawn from the study of the modular organization.

The study of the modular structure can also explain why the average
weight of the attraction basins of critical networks vanishes in the
infinite size limit.

Thus it would be interesting to investigate analytically the distribution of
size and number of the modules. We think that the approach that we
used to estimate the number of relevant elements, consisting in adding
a new element to the network and computing the modification of the
modular structure, could be a good tool for a more quantitative
understanding of the modular organization of critical networks, if the
approximation that we used can be in some way controlled.

\vspace{0.5cm}
The modular organization of Kauffman networks is a consequence of the
finite connectivity of this model. It is because of the finite
connectivity that some elements become stable, and act as a barrier in
the transmission of information among different modules.
Thus the structures described in this work are peculiar to this kind of
cellular automata, and they are not common to other models, for other aspects
very similar to this one, such as asymmetric neural networks.
% \cite{BP4}.
It would be interesting to understand if and how the existence of a
modular organization is related to the biological situation that Kauffman
networks are intended to model, {\it i.e.} genetic regulatory systems.

We think that the spontaneous emergence of a modular organization in critical
networks is one of the most interesting features of such systems, and could
have played an important role in the emergence of cellular organization from
an hypothetical network of chemical reactions where the primordial regulatory
rules were of a probabilistic type. One of the advantages of modularity is
the possibility to build many independent objects, which could have
been shaped more
easily by natural selection. This interpretation is highly speculative
but very appealing, and raises for the biological modelization the
challenging question about the mechanisms which could drive to criticality
such an hypothetical ensemble of self regulated chemical reactions.

We argue from this study that for large system size the typical scale of the
number and of the length of the attractors increase much faster than
$\sqrt N$, as it was observed by Kauffman in early works \cite{K69}.
The $\sqrt N$ law was considered by Kauffman an important argument
in favor of the fact that randomly assembled networks of regulatory systems
can behave in a way very reminiscent of the behavior of true biological
organisms. This law, and thus the biological analogy, still holds
approximately for a small enough number of elements. When the number
of genes becomes large, on the other hand, the number of cycles (in the
biological metaphor, the number of cellular types) and especially their
length (which in the biological metaphor represents cycle cell time) become
too large to allow a biological interpretation. In this case
it seems unlikely that critical random boolean
networks behave as control networks in the cells, and that the
scaling laws observed for real organisms are just a reflex of the typical
properties of random regulatory networks. Random boolean networks with
a number of elements of the same order as the number of genes in
higher eukariots have with finite  probability too many and too long
attractors, and some kind of selection must come into play to
maintain the approximate power law scaling observed for small
systems. But it may be that genetically interesting networks are not 
vanishingly rare in the ensemble of critical Kauffman networks
in the infinite size limit. In the biological picture proposed by
Kauffman, this would mean that some properties of real cells can still
be interpreted as typical properties of an ensemble of randomly assembled
regulatory networks, even if this ensemble has not to be identified
with the whole ensemble of critical Kauffman networks.

\section*{Acknowledgments}
U.B. would like to thank Henrik Flyvbjerg and Peter Grassberger for
interesting discussions and for reading the manuscripts.

\end{document}